\documentclass[12pt]{article}
\usepackage{float}
\usepackage[caption = false]{subfig}
\usepackage[final]{graphicx}
\usepackage{latexsym,graphicx,multirow}
\usepackage{amssymb}
\usepackage{amsmath}
\usepackage{amscd}
\usepackage{amsthm}
\graphicspath{{images/}}
\usepackage[left=2cm,top=2.5cm,right=2.5cm,bottom=1.5cm]{geometry}
\usepackage{hyperref}
\usepackage{color}

\usepackage{epstopdf}
\usepackage{cite}
\usepackage{float}
\usepackage[utf8]{inputenc}

\begin{document}

    \begin{center}
        \large{\bf{Holographic dark energy through Kaniadakis entropy in non flat universe}} \\
        \vspace{10mm}
   \normalsize{ Suresh Kumar P$^{a,1}$, Bramha Dutta Pandey$^{b,2}$, Umesh Kumar Sharma $^{c,3}$, Pankaj$^{d,4}$} \\
    \vspace{5mm}
    \normalsize{$^{a,\:b,\:c}$ Department of Mathematics, Institute of Applied Sciences and Humanities, GLA University
        Mathura-281406, Uttar Pradesh, India}\\
    \vspace{2mm}
    \normalsize{$^{a,\:b}$ IT Department (Math Section), University of Technology and Applied Sciences-Salalah, Oman}\\
    \vspace{2mm}
    \normalsize{$^d$ Department of Mathematics(BSH), Pranveer Singh Institute Of Technology, Kanpur-209305, Uttar Pradesh, India}\\
    \vspace{2mm}

    $^1$E-mail: sureshharinirahav@gmail.com\\
    $^2$E-mail:bdpandey05@gmail.com\\
    $^3$E-mail: sharma.umesh@gla.ac.in \\
    $^4$E-mail: pankaj.fellow@yahoo.co.in \\
       \vspace{10mm}

\end{center}

\begin{abstract}
By extending the standard holographic principle to a cosmological framework and combining the non-flat condition with the Kaniadakis entropy, we construct the non-flat  Kaniadakis holographic dark energy model. The model employs Kaniadakis parameter $K$ and a parameter $c$. Derivation of the differential equation for KHDE density parameter to describe the evolutionary behavior of the universe is obtained. Such a differential equation could explain both the open as well as closed universe models. The classification based on  matter and dark energy (DE) dominated regimes show that the KHDE scenario may be used to specify the Universe's thermal history and that a quintom regime can be encountered. For open and closed both the cases, we find the  expressions for the deceleration parameter and the equation of state (EoS) parameter. Also, by varying the associated parameters, classical stability of the method is established. On considering the curvature to be positive, the universe favors the quintom behavior for substantially smaller values as opposed to the flat condition, when only quintessence is attained for such $K$ values. Additionally, we see a similar behavior while considering the curvature to be negative for such $K$ values. Therefore, adding a little bit of spatial geometry that isn't flat to the KHDE enhances the phenomenology while maintaining $K$ values at lower levels. To validate the model parameters, the most recent $30\;H(z)$ dataset measurements, in the redshift range $0.07 \leq z \leq 1.965$ are utilized. In addition, the distance modulus measurement from the current Union 2.1 data set of type Ia supernovae are employed. \\

\smallskip


\end{abstract}

\section{\textbf{Introduction}}
The majority view in contemporary cosmology holds that both the early and late universes witnessed accelerated expansion, which is backed by a vast number of cosmic observations \cite{Riess98,Perl99}. The first strategy is to continue using general relativity to explain gravity theory while introducing new types of matter, such as inflation \cite{Olive89,Bartolo04} or the idea of dark energy \cite{Copeland06,Cai09}. The second route involves creating extended and modified gravitational theories, which, while they generally provide an additional degree of freedom that might result in acceleration, nevertheless have general relativity as a specific limit \cite{Nojiri06,Capozziello11,Cai15}. Mathematically, the Universe's acceleration can alternatively be described through holographic dark energy(HDE) \cite{Li04,Wang17}  and holographic inflation \cite{Nojiri19} in an excellent way. But, technically speaking, it does not fit in the above two solution approaches.
However, the cosmic application \cite{Fischler98,Bak99,Horava2000} of the holographic principle \cite{tHooft93,Susskind94,Bousso02} allows for a different explanation of how dark energy was created.
The concept utilizes thermodynamics of black holes and the relationship between a quantum field theory's Ultraviolet cutoff and  largest theory's distance. A quantum field theory must have an ultraviolet cutoff for it to be applicable across large distances \cite{Cohen98}. In particular, to prevent the system from collapsing into a black hole, the total energy in a particular system with volume-dependent entropy can not be more than an equivalent-sized black hole's mass whose area depends on entropy. When the entire universe is seen as a system, holographic vacuum energy, or HDE with a dynamical aspect, may be extracted \cite{Li04, Wang17}.\\

HDE's cosmic effects turn out to be both fascinating \cite{Li04,Wang17,Horvat04,Pavon05,Huang04,Nojiri05,Setare06} and consistent with observations \cite{Zhang05,Li09,Feng07,Zhang09,Lu09,Micheletti09}. The fundamental expression in the development of HDE is the one that relates a system's entropy to  its radius. The Boltzmann-Gibbs entropy has been applied to black holes and other cosmic phenomena to produce the most famous Bekenstein-Hawking entropy. In contrast, the Boltzmann-Gibbs entropy's one-parameter generalization was proposed by Kaniadakis \cite{Kaniadakis02,Kaniadakis05}. This is the result of a self-consistent and coherent relativistic statistical theory which maintains the basic guidelines of normal statistical theory. Continuous deformation of the original Maxwell-Boltzmann distribution function by one parameter is made in such an expanded statistical theory.
Thus, as a limited scenario, the usual statistical theory is restored. \\
The holographic approach is used to produce HDE using the black hole entropy expression. As a result, many versions of the theory may be generated by varying the entropy. Applying Kaniadakis entropy in a black-hole framework leads to 
\begin{equation}
S_K=\frac{1}{K} \sinh(K S_{BH}), \label{e1}
\end{equation}
where $S_{BH}$ is the Bekenstein-Hawking entropy. Bekenstein-Hawking entropy is the limiting case of(\ref{e1}) i.e. $\lim_{K\rightarrow 0} S_K=S_{BH}$. $K<<1$ (especially $-1<K<1$) is expected for (\ref{e1}) to recover standard Bekenstein-Hawking value. The basic Kaniadakis entropy can therefore be extended for small $K$, yielding \cite{Drepanou21}
\begin{equation}
S_K = S_{BH} + \frac{S_{BH}^3}{6} + \mathrm{O}(K^4).   \label{e2}
\end{equation}
Various cosmological aspects of KHDE model, theoretically as well as observationally in \cite{Hernandez-Almada:2021rjs,Hernandez-Almada21, Moradpour20, Jawad21, Nojiri22,  Drepanou21, dubey2022some, singh2022statefinder, Sadeghi:2022fow, P:2022csj,Ghaffari:2021xja,Rani:2022upi,Abreu:2017hiy}.\\

The combined study of the Planck Collaboration's anisotropic power spectra of the CMB and luminosity distance data indicates a non-flat universe at a $99\%$ confidence level \cite{DiValentino20}. Despite being much less than other energy components, it is still conceivable that the spatial curvature makes a contribution to the Friedmann equation. Studying a universe with a spatial curvature that is only tangentially permitted by the inflation hypothesis, along with observations, is therefore of more than just academic interest. In literature, Guo \cite{Huang04} generalized the HDE model of M. Li \cite{Li04, Wang17} by insisting non-flatness into consideration. It was concluded that in order to maintain thermodynamics' second law into picture, phantom-like situations will not exist. Setare \cite{Setare06} studied the HDE model by considering the interaction among the constituent sectors in a non flat universe. Sharma et al.  \cite{Sharma21} studied the KHDE model with an apparent horizon in a non flat universe. Recently, Adhikari \cite{Adhikary21} studied the Barrow HDE model with the largest theory's distance as a future event horizon to analyze the behavior of various cosmic parameters. The fractional energy density due to curvature is taken as $\Omega_{k_{sc}}$. $\Omega_{k_{sc}}>0$ signifies the spatially closed whereas $\Omega_{k_{sc}}<0$ corresponds to a spatially open universe. The Planck 2018 CMB temperature and polarization anisotropy data favors the non flatness of the universe with  $\Omega_{k_{sc}}=-0.044_{-0.015}^{+0.018}$\cite{Planck18}. Additionally, by combining $P18$ data with the full-shape (FS) universe power spectrum and adding the $\Omega_{k_{sc}} = 0.0023\pm 0.0028$ obtained from this collaboration  together with calculations from the BOSS DR12 CMASS example, one can attain results with weaker constraints \cite{Vagnozzi20}. In any scenario, a considerable number of scholars have been motivated to put tenable constraints on $\Omega_{k_{sc}}$ because the evolution of the universe is under the influence of spatial curvature \cite{DiValentino20}. In light of the above discussions, our current study focuses on constructing and analyzing  the non-flat universe with Kaniadakis holographic dark energy with future event horizon as IR cutoff.
\\

The manuscript starts with a brief introduction. Next section is devoted to formulating KHDE to describe the events of the closed and open Friedmann-Robertson-Walker metric. We continue our extensive analysis of the cosmic behavior in Section III by concentrating towards  dark energy density, EoS and deceleration parameters. Additionally, we'll examine the squared sound speed to confirm the model's traditional stability. In Section IV, the techniques used to analyze the data for this study are described.  Last section is dedicated to the summary of obtained results and conclusions.

\section{Kaniadakis Holographic Dark Energy (KHDE) in Non-Flat Geometry}

In this section, by assuming that spatial curvature is non-zero, we seek to generate holographic dark energy.We pay special attention to a non-flat FRW line element. Such a line element with $a(t)$ as scale factor  is expressed by
\begin{equation}
\mathrm{d}s^2=a(t)^2\left[\dfrac{\mathrm{d}r^2}{1-k_{sc}r^2}+r^2\mathrm{d}\Omega^2\right]-\mathrm{d}t^2, \label{e3}
\end{equation}
where the choice  $k_{sc}=-1,0,+1$ correspondingly result in open, flat, and closed spatial curvatures.\\

Using (\ref{e2}) and the inequality $\rho_d L^4\leq S$ will give the expression for the KHDE density $\rho_d$ given by \cite{Drepanou21, Pandey22}
\begin{equation}
\rho_d=\frac{3c^2M_p^2}{L^2}+K^2M_p^6L^2, \label{e4}
\end{equation}
where $L, c$ and $K$ are length of horizon, model and Kaniadakis parameters respectively and $M_p$ is the reduced Plank's mass given by $M_p=\dfrac{1}{\sqrt{8\pi G}}$. The standard Bakenstein-Hawking entropy is recovered from Kaniadakis entropy in equation (\ref{e4}) for $K=0$.\\

To derive the basic results, we have considered the presence of only two sectors in the universe namely: dark energy and dark matter. If $\rho_m$ represents the dark matter density and $p_d$ the KHDE pressure,  the corresponding Friedmann equations are given by 
\begin{equation}
3H^2+3\dfrac{k_{sc}}{a^2}=\rho_d+\rho_m,  \label{e5}
\end{equation} 
\begin{equation}
2\dot{H}+3H^2+\dfrac{k_{sc}}{a^2}=-p_d,  \label{e6}
\end{equation}
with Hubble parameter $H$ given by $H=\dfrac{\dot{a}}{a}$. The conservation equation for the dark matter and KHDE sectors are given by
\begin{equation}
\dot{\rho_m}+3H\rho_m=0, \label{e7}
\end{equation}
\begin{equation}
\dot{\rho_d}+3H(1+w_d)\rho_d=0, \label{e8}
\end{equation}
with KHDE's equation of state (EoS) parameter is $w_d$ defined by $w_d=\dfrac{p_d}{\rho_d}$. For the sake of further analysis, the dark matter, KHDE and curvature density parameters are defined respectively by $\Omega_m=\dfrac{\rho_m}{3M_p^2H^2},\; \Omega_d=\dfrac{\rho_d}{3M_p^2H^2}$ and $\Omega_{k_{sc}}=\dfrac{k_{sc}}{a^2H^2}$. In the light of equation(\ref{e5}), we get the relation among density parameters as
\begin{equation}
\Omega_d+\Omega_m=1+\Omega_{k_{sc}}. \label{e5a}
\end{equation}
The next stage is to define the theory's longest length $L$, specifically the holographic horizon, which is a part of the description of HDE. In the case of the flat-universe, future event horizon $R_h$ will serve the purpose of the theory's largest distance $L$ to describe the accelerated expansion. But for the non-flat universe model, a suitable modification is needed to get an appreciable behavior of cosmic parameters for the accelerated expanding universe\cite{Huang04, Setare06}. The suitably modified $L$ will be discussed in subsections concerning closed and open spatial curvatures.
\subsection{Case I: Closed Spatial Curvature ($k_{sc}=1$)}
For the case of closed spatial curvature, the largest theory's distance $L$ is defined by $L=ar(t)$ \cite{Huang04, Setare06} with $r(t)$ defined by the relation 
\begin{equation}
\int_{0}^{r(t)} \dfrac{1}{\sqrt{1-k_{sc}r^2}}\mathrm{d}r=\int_{t}^{\infty} \dfrac{\mathrm{d}t}{a}=\dfrac{R_h}{a}, \label{e9}
\end{equation}
which gives 
\begin{equation}
r(t)=\dfrac{1}{\sqrt{k_{sc}}} \sin{y}, \label{e10}
\end{equation}
where 
\begin{equation}
y=\sqrt{k_{sc}}\dfrac{R_h}{a}=\sqrt{k_{sc}}\int_{x}^{\infty}\dfrac{\mathrm{d}x}{Ha}, \label{e11}
\end{equation}
where $a=e^x$. 
We are interested to analyze the behavior of various density parameters and hence will use expressions in terms of their present value as
\begin{equation}
\Omega_m=\dfrac{\Omega_{m_0}H_0^2}{a^3H^2},\;\;\Omega_{k_{sc}}=\dfrac{\Omega_{{k_{sc}}_0}H_0^2}{a^2H^2}, \label{e13}
\end{equation}
by considering 
\begin{equation}
\dfrac{\Omega_{k_{sc}}}{\Omega_m}=a\gamma, \label{e14}
\end{equation}
where $\gamma=\dfrac{\Omega_{{k_{sc}}_0}}{\Omega_{m_0}}$.\\

The universe's curvature is related to the energy density and expansion rate using the Friedmann equation (\ref{e5a}) , given by \cite{Huang04}
\begin{equation}
H=\dfrac{H_0\sqrt{\Omega_{m_0}}\sqrt{a^{-1}-\gamma}}{a\sqrt{1-\Omega_d}}, \label{e15}
\end{equation}
Using equations (\ref{e10}), (\ref{e11}) and (\ref{e15}), we get
\begin{equation}
L=\dfrac{a}{\sqrt{k_{sc}}}\sin\left[ \sqrt{k_{sc}}\int_{x}^{\infty} \dfrac{1}{\sqrt{\Omega_{m_0}}H_0}\sqrt{\left(\dfrac{1-\Omega_d}{a^{-1}-\gamma}\right)} \mathrm{d}x\right]. \label{e16}
\end{equation}

Another expression for $L$ is obtained through equations (\ref{e3},\ref{e4}) and (\ref{e15}) given by
\begin{equation}
L=\sqrt{\dfrac{3H^2\Omega_d-\sqrt{9H^4\Omega_d^2-12c^2 K^2 M_p^4}}{2 K^2 M_p^4}}. \label{e17}
\end{equation}
As both the equations (\ref{e16}) and(\ref{e17}) are giving the value of $L$, we get
\begin{equation}
\dfrac{a}{\sqrt{k_{sc}}}\sin\left[ \sqrt{k_{sc}}\int_{x}^{\infty} \dfrac{1}{\sqrt{\Omega_{m_0}}H_0}\sqrt{\left(\dfrac{1-\Omega_d}{a^{-1}-\gamma}\right)} \mathrm{d}x\right]=\sqrt{\dfrac{3H^2\Omega_d-\sqrt{9H^4\Omega_d^2-12c^2K^2M_p^4}}{2K^2M_p^4}}.    \label{e18}
\end{equation}
Using the transformation $a=e^x$ in  (\ref{e18}) and differentiating w.r.t. $x$ we get
\begin{equation}
\Omega_d'=2\Omega_d(1-\Omega_d)\left(\dfrac{B\sqrt{\Omega_d}}{A^{3/2}}\sqrt{\dfrac{A+B}{2c^2}-3k_{sc}e^{-2x}}-\dfrac{A+B}{A}+\dfrac{5-4e^x\gamma}{2-2e^x\gamma}\right),  \label{e19}
\end{equation}
where $A=\dfrac{3e^{-3x}(1-e^x\gamma)H_0^2\Omega_{m_0}\Omega_d}{1-\Omega_d}$, $B=\sqrt{A^2-12c^2K^2M_p^4}$.\\

The differential equation (\ref{e19}) describes the evolution of the KHDE by supposing a closed universe with dark matter and dark energy as its primary components. The case of $k_{sc}=0$ i.e. the curvature density $\Omega_{k_{sc}}$ to be $0$ $ (\gamma=0) $ corresponds to KHDE in flat universe \cite{Drepanou21}. When $ K\rightarrow 0 $, we get $ B=A $, leads the equation (\ref{e19}) to standard HDE model of non-flat universe's differential equation for the DE density \cite{Huang04},  $\dfrac{\Omega_d'}{\Omega_d^2} = (1-\Omega_d) \left(\dfrac{1}{(1-a \gamma)\Omega_d}+\dfrac{2 \cos y}{c\sqrt{\Omega_d}} \right)$.
We point out that when $K\rightarrow 0$ and $\gamma=0$, it leads (\ref{e19}) in retrieving the equivalent differential equation of the conventional HDE model for the flat universe. This equation is $\Omega_d' = \Omega_d(1-\Omega_d) \left(1 + 2\sqrt{\dfrac{3M_p^2\Omega_d}{3c^2M_p^2}}\right)$, and being independent of $x$, it can have an implicit analytic solution \cite{Li04, Wang17
}.\\

We shall now examine how the KHDE's EoS parameter $w_d= p_d /\rho_d$ behaves. Because of the conservation of matter sector, the Friedmann equations (\ref{e5}-\ref{e6}) forces the conservation of dark energy sector, i.e.
\begin{equation}
\dot{\rho_d}+3H\rho_d(1+w_d)=0, \label{e20}.
\end{equation}
Differentiating (\ref{e4}), using (\ref{e10}), (\ref{e11}) with $\dot{L}=LH-\cos{y}$ and inserting into (\ref{e8}), we get

\begin{equation}
w_d=-1+\dfrac{2B}{3A}\left(1-\dfrac{1}{c}\sqrt{\Omega_d}\sqrt{\dfrac{A+B}{2A}}\cos{\left[ce^{-x}\sqrt{\dfrac{6 k_{sc}}{A+B}}\right]}\right). \label{e24}
\end{equation}

Hence Obtaining $w_d$ requires knowledge of $\Omega_d$ from  (\ref{e19}). On considering $k_{sc}=0$, we achieve the EoS parameter expression obtained for flat-universe \cite{Drepanou21}. When  $ K=0 $ equation (\ref{e24}) becomes the EoS equation for non flat universe $w_d=-\dfrac{1}{3}-\dfrac{2}{3c}\sqrt{\Omega_d} \cos y$ . For $ k_{sc}=K=0 $, the relation (\ref{e24}) leads to the standard holographic dark energy scenario $w_d=-\dfrac{1}{3}-\dfrac{2}{3c}\sqrt{\Omega_d}$ \cite{Li04, Wang17}.

From the Friedmann second equation (\ref{e6}) and Hubble parameter (\ref{e15}), the deceleration parameter $ q=-1-\dfrac{\dot{H}}{H^2} $ is deduced as 

\begin{equation}
q=\dfrac{1}{2}\left(1+3w_d\Omega_d+k_{sc}\dfrac{1-\Omega_d}{H_0^2\Omega_{m_0}\left(e^{-x}-\gamma\right)}\right). \label{e24a}
\end{equation}

\subsection{Case II: Open Spatial Curvature ($k_{sc}=-1$)}
In this subsection the analysis is based on the negative spatial curvature consideration, i.e. $k_{sc}=-1$. The largest theory's distance $L$ is given by $L=ar(t)$, where the expression for $r(t)$ can be obtained using the relation  
\begin{equation}
\int_{0}^{r(t)} \dfrac{1}{\sqrt{1+k_{sc}r^2}}\mathrm{d}r=\dfrac{R_h}{a}, \label{e25}
\end{equation}
which gives the relation for $r(t)$ as
\begin{equation}
r(t)=\dfrac{1}{\sqrt{|k_{sc}|}}\sinh{y}, \label{e26}
\end{equation}
where, using the transformation $a=e^x$ we get
\begin{equation}
y=\sqrt{|k_{sc}|}\dfrac{R_h}{a}=\sqrt{|k_{sc}|}\int_{x}^{\infty} \dfrac{1}{Ha}\mathrm{d}x. \label{e27}
\end{equation}
The calculation for $L$ will be performed similar to the closed spatial curvature case and we get the relation
\begin{equation}
\dfrac{a}{\sqrt{|k_{sc}|}}\sinh\left[ \sqrt{|k_{sc}|}\int_{x}^{\infty} \dfrac{1}{\sqrt{\Omega_{m_0}}H_0}\sqrt{\left(\dfrac{1-\Omega_d}{a^{-1}-\gamma}\right)} \mathrm{d}x\right]=\sqrt{\dfrac{3H^2\Omega_d-\sqrt{9H^4\Omega_d^2-12c^2 K^2 M_p^4}}{2 K^2 M_p^4}}.    \label{e28}
\end{equation}
Taking the $x-$ derivative of the relation (\ref{e28}) by using the transformation $a=e^x$ and applying equation (\ref{e15}) leads to the differential equation 
\begin{equation}
\Omega_d'=2\Omega_d(1-\Omega_d)\left(\dfrac{B\sqrt{\Omega_d}}{A^{3/2}}\sqrt{\dfrac{A+B}{2c^2}+3|k_{sc}| e^{-2x}}-\dfrac{A+B}{A}+\dfrac{5-4e^x\gamma}{2-2e^x\gamma}\right),  \label{e29}
\end{equation}
where $A=\dfrac{3e^{-3x}(1-e^x\gamma)H_0^2\Omega_{m_0}\Omega_d}{1-\Omega_d}$, $B=\sqrt{A^2-12c^2 K^2 M_p^4}$.

The development of KHDE for dust matter in an open universe is given by the differential equation (\ref{e29}). The case of $k_{sc}=0$ leads to the KHDE for flat universe \cite{Drepanou21, Pandey22}. While as $K=0$, the current model tend towards the standard holographic dark energy scenario for open universe \cite{Huang04}, $\dfrac{\Omega_d'}{\Omega_d^2} = (1-\Omega_d) \left(\dfrac{1}{(1-a \gamma)\Omega_d}+\dfrac{2 \cosh y}{c\sqrt{\Omega_d}}  \right)$, using the relations $ \cosh^2 y-\sinh^2y=1  $ and $ L=c\sqrt{\dfrac{3}{A}}=\dfrac{c}{H\sqrt{\Omega_d}} $. If we restrict both, i.e. $k_{sc}=0$ and  $K\rightarrow 0$,  it returns the usual HDE in a flat universe  $\Omega_d' = \Omega_d(1-\Omega_d) \left(1 + 2\sqrt{\dfrac{3M_p^2\Omega_d}{3c^2M_p^2}}\right) $  \cite{Li04, Wang17}. \\

Now we will obtain the EoS parameter expression for open spatial curvature in a way exactly similar to the closed universe case. Taking the time derivative of (\ref{e4}), using equations (\ref{e8}), (\ref{e26}) and (\ref{e27}) we get 
\begin{equation}
w_d=-1+\dfrac{2B}{3A}\left(1-\dfrac{1}{c}\sqrt{\Omega_d}\sqrt{\dfrac{A+B}{2A}}\cosh{\left[ce^{-x}\sqrt{\dfrac{6|k_{sc}|}{A+B}}\right]}\right). \label{30}
\end{equation}
On considering $k_{sc}=0$, the EoS parameter expression for KHDE model with flat universe is obtained \cite{Drepanou21, Pandey22}. $K\rightarrow 0$ leads to the EoS parameter expression for usual HDE model with open spatial curvature $w_d=-\dfrac{1}{3}\left(1+\dfrac{2}{c}\sqrt{\Omega_d} \cosh y \right)$  . Standard holographic dark energy model \cite{Li04, Wang17} is recovered as a limiting case of the current model, i.e. for $k_{sc}=0$ and $K\rightarrow 0$, we get  $w_d=-\dfrac{1}{3}\left(1+\dfrac{2}{c}\sqrt{\Omega_d} \right)$.
The deceleration parameter expression is provided using the same formula as in the closed case. 
\begin{equation}
q=-1-\dfrac{\dot{H}}{H^2}=\dfrac{1}{2}\left(1+3w_d\Omega_d+k_{sc}\dfrac{1-\Omega_d}{H_0^2\Omega_{m_0}\left(e^{-x}-\gamma\right)}\right). \label{e30a}
\end{equation}

In order to examine the KHDE model's conventional stability, the squared sound speed is stated as 
\begin{equation}
v_s^2=\dfrac{\mathrm{d}p_d}{\mathrm{d}\rho_d}=\dfrac{\rho_d}{\dot{\rho_d}}\dot{w_d}+w_d.  \label{e31}
\end{equation}
\section{Cosmological Evolution}
The KHDE model was developed in the part before this one with a future event horizon by considering the spatial geometry to be open as well as closed. Various equations pertaining to the evolutionary behavior of the universe such as KHDE density, EoS and deceleration parameters were obtained. As a result, we may now conduct a thorough analysis of the resultant cosmological behavior. The equations (\ref{e19}) and (\ref{e29}) can be solved analytically only for $\gamma=0$ and $K=0$. We must seek a  numerical solution for the general case. Using the straightforward equation $x =\ln {a} = -\ln{(1 + z)}$, we may predict its behavior in terms of the redshift $z$  as long as the solution for $\Omega_d$ is available. Kaniadakis entropy, being an even function, restricts $K$ values to be positive only.\\

\begin{figure}[H]
\subfloat[]{\includegraphics[scale=0.4]{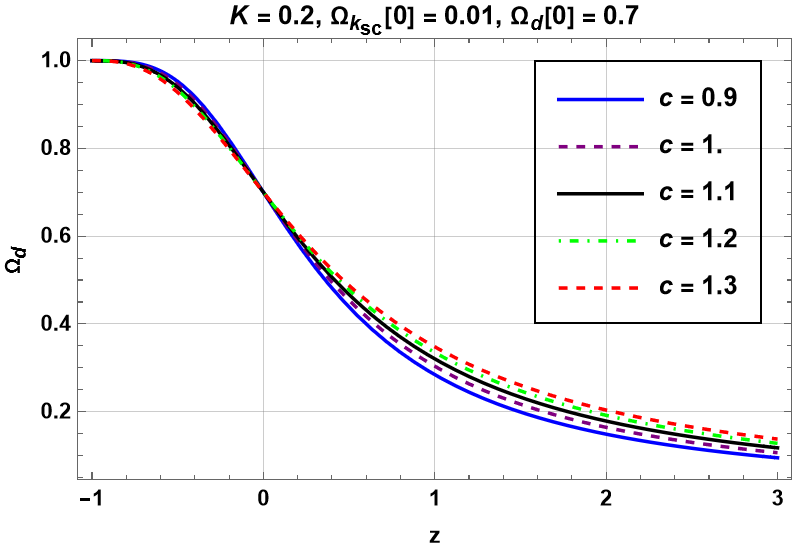}}
\subfloat[]{\includegraphics[scale=0.4]{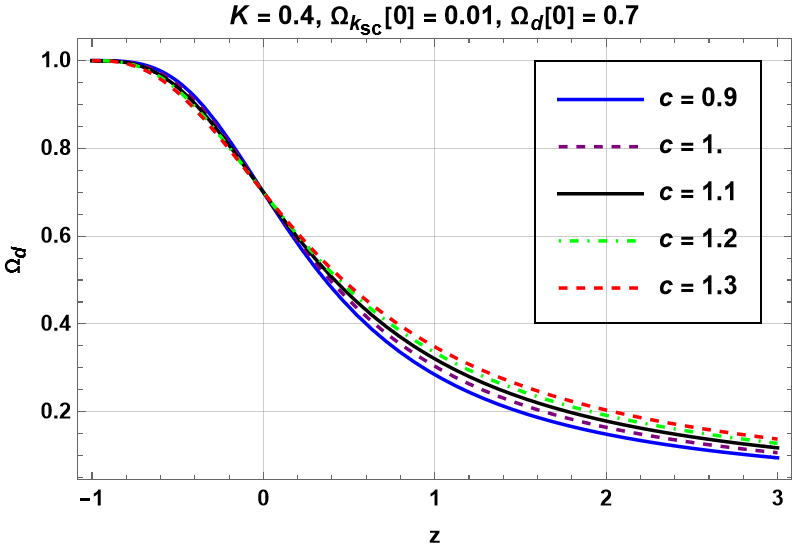}}
\subfloat[]{\includegraphics[scale=0.4]{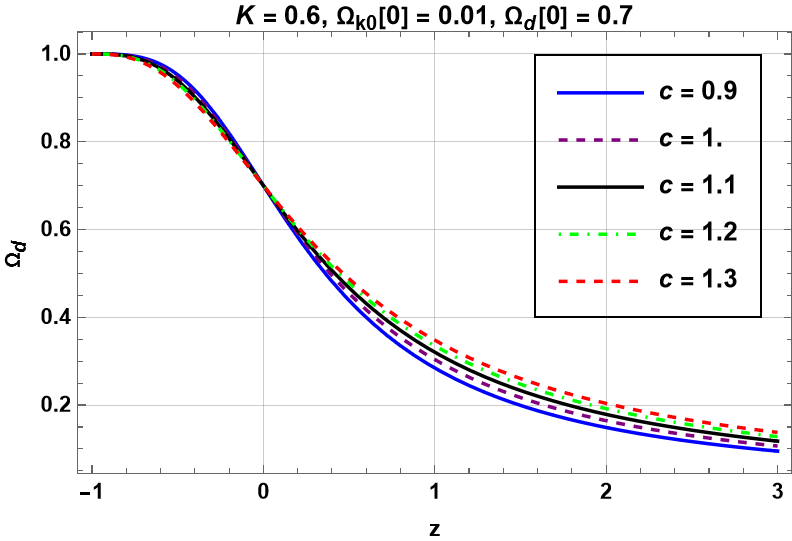}}
\caption{\begin{small}The variation in the KHDE density parameter $\Omega_d$ with redshift $z$ for the closed universe,  in units when  $M_p^2=1, \quad H_0=67.9$. In order to be consistent with data, we now imposed $\Omega_d(x =- \ln(1 + z) = 0)= \Omega_d[0]= 0.7$.\end{small}}
\label{P1}
\end{figure}

\begin{figure}[H]
\subfloat[]{\includegraphics[scale=0.4]{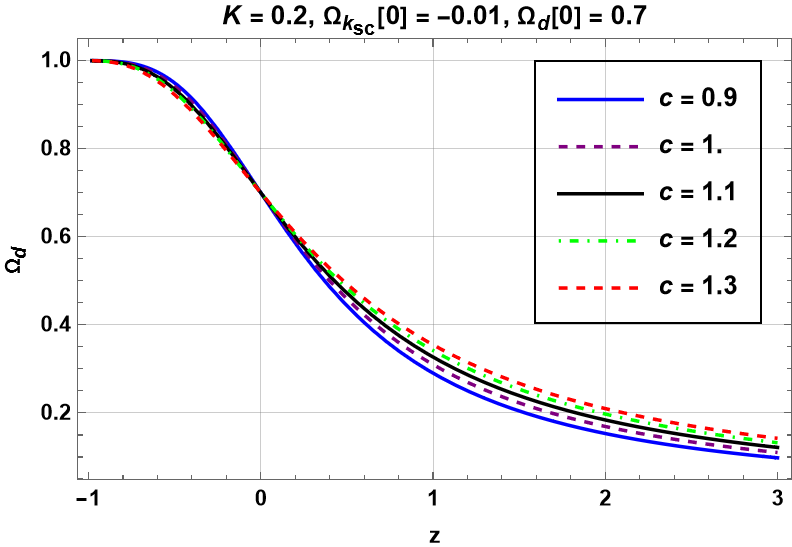}} 
\subfloat[]{\includegraphics[scale=0.4]{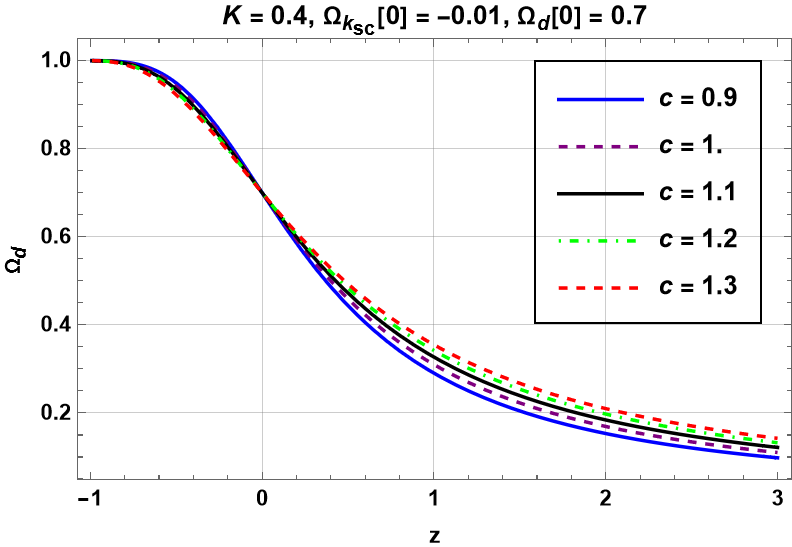}}
\subfloat[]{\includegraphics[scale=0.4]{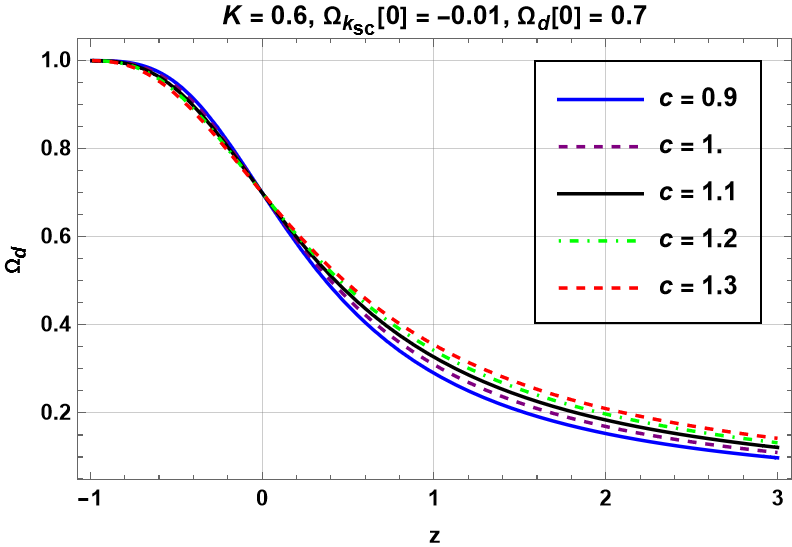}}
\caption{\begin{small}	The variation in the KHDE density parameter $\Omega_d$ with redshift $z$ for the open universe,  in units when  $M_p^2=1, \quad H_0=67.9$. In order to be consistent with data, we now imposed $\Omega_d(x =- \ln(1 + z) = 0)= \Omega_d[0]= 0.7$.\end{small}}
\label{P2}
\end{figure}

Inline with observational data, equations (\ref{e19}) and (\ref{e29}) are solved numerically by considering $\Omega_d(x=-\ln{(1+z)}=0)=0.7\equiv \Omega_{d_0}$, $\Omega_m(x=-\ln{(1+z)}=0)=0.29\equiv \Omega_{m_0},$ and $\Omega_k(x=-\ln{(1+z)}=0)=0.01\equiv \Omega_{k_0}$. By varying $c$ and fixing $K$, figures \ref{P1} and \ref{P2} are plotted.
Figure \ref{P1} explains the closed universe's thermal history whereas  figure \ref{P2} characterizes the open universe's thermal history. Both the figures  \ref{P1} and \ref{P2} imply that dark matter ruled the whole universe in the past, partially KHDE dominated at present and fully KHDE dominated in the near or far future.

\begin{figure}[H]
\subfloat[]{\includegraphics[scale=0.4]{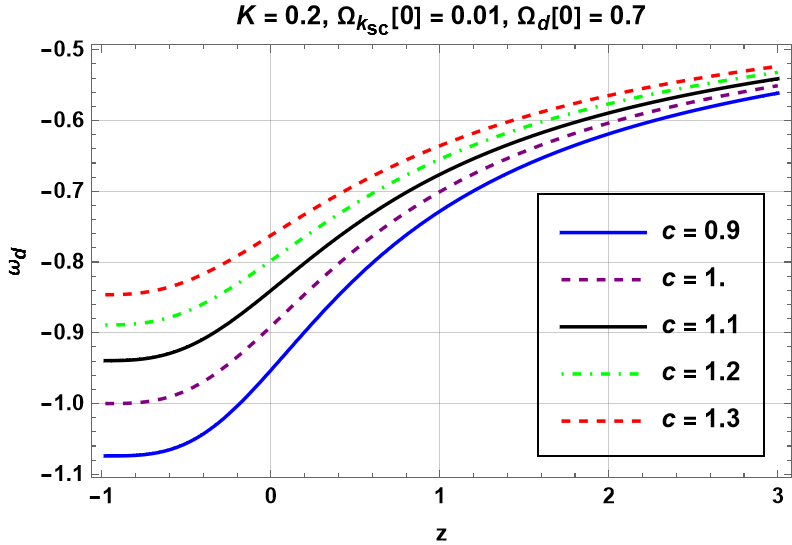}} 
\subfloat[]{\includegraphics[scale=0.4]{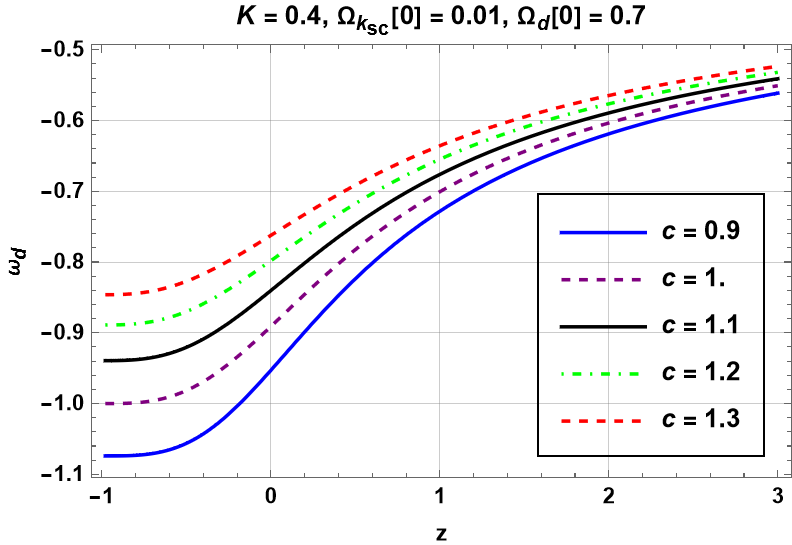}}
\subfloat[]{\includegraphics[scale=0.4]{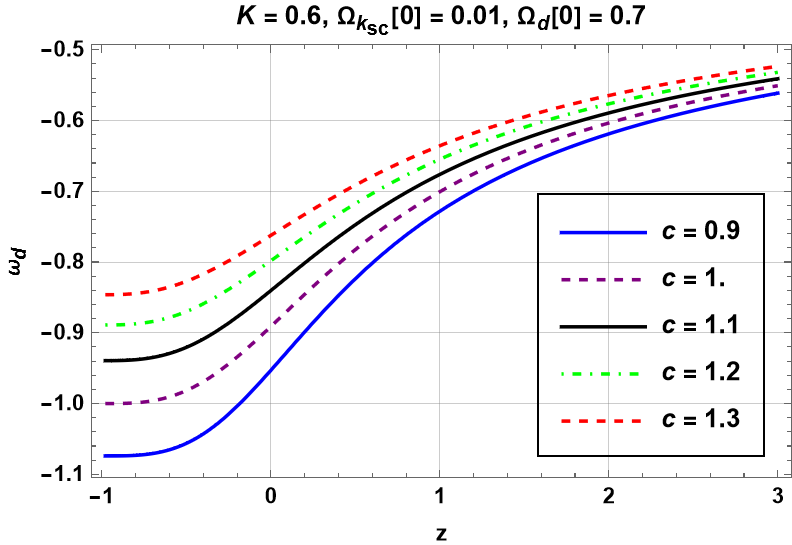}}
\caption{\begin{small}The variation in the EoS  parameter $w_d$ with redshift $z$ for the closed universe,  in units when  $M_p^2=1, \quad H_0=67.9$. In order to be consistent with data, we now imposed $\Omega_d(x =- \ln(1 + z) = 0)= \Omega_d[0]= 0.7$.\end{small}}
\label{P3}
\end{figure}

\begin{figure}[H]
\subfloat[]{\includegraphics[scale=0.4]{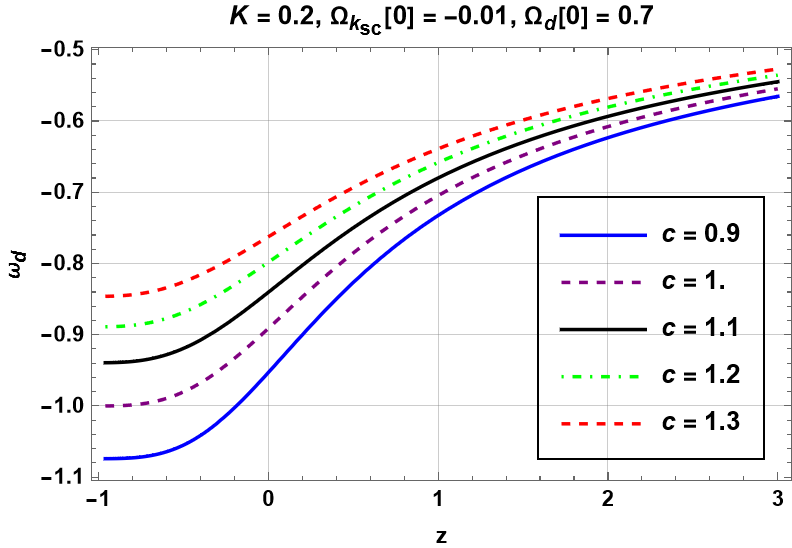}} 
\subfloat[]{\includegraphics[scale=0.4]{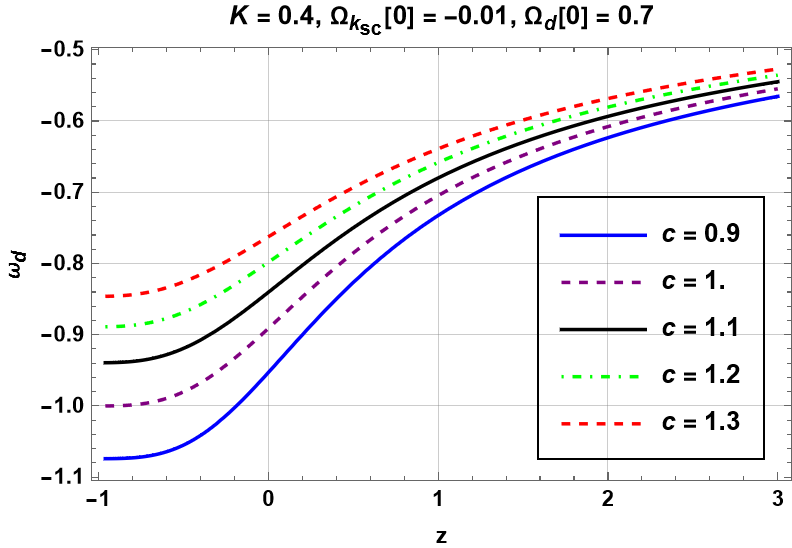}}
\subfloat[]{\includegraphics[scale=0.4]{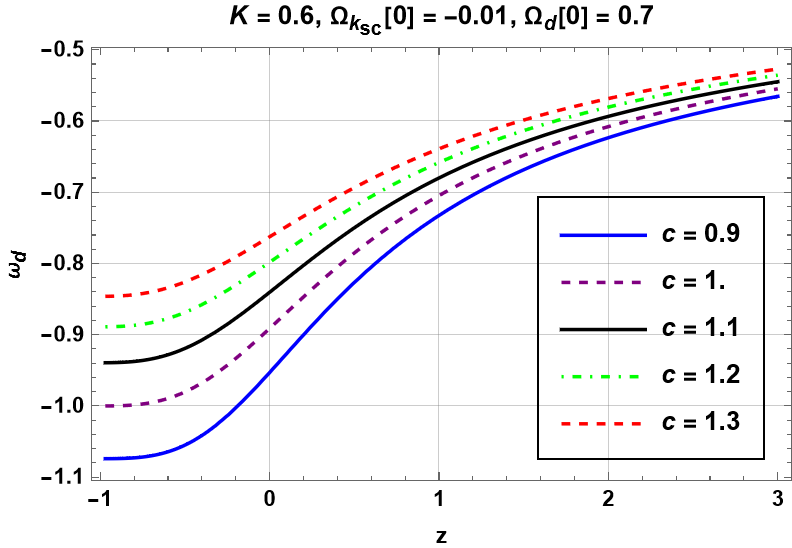}}
\caption{\begin{small}	The variation in the EoS  parameter $w_d$ with redshift $z$ for the open universe,  in units when  $M_p^2=1, \quad H_0=67.9$. In order to be consistent with data, we now imposed $\Omega_d(x =- \ln(1 + z) = 0)= \Omega_d[0]= 0.7$.\end{small}}
\label{P4}
\end{figure}

The EoS parameter is plotted against redshift $z$ in figure \ref{P3} for closed universe model and in \ref{P4} for open universe by assuming $K$ to be fixed and varying $c$ . The KHDE model never crosses the line $w_d=-1$, shows pure quintessence behavior for $c<1$, and $c=1$ results into the $\Lambda-$ CDM model. For $c$ values greater than $1$, the EoS trajectories evolve in the quintessence region, and by going across the divide line $w_d=-1$, in the near or far future enters the phantom zone. As a result, when $c>1$, the KHDE model acts like quintom.\\

\begin{figure}[H]
\subfloat[]{\includegraphics[scale=0.4]{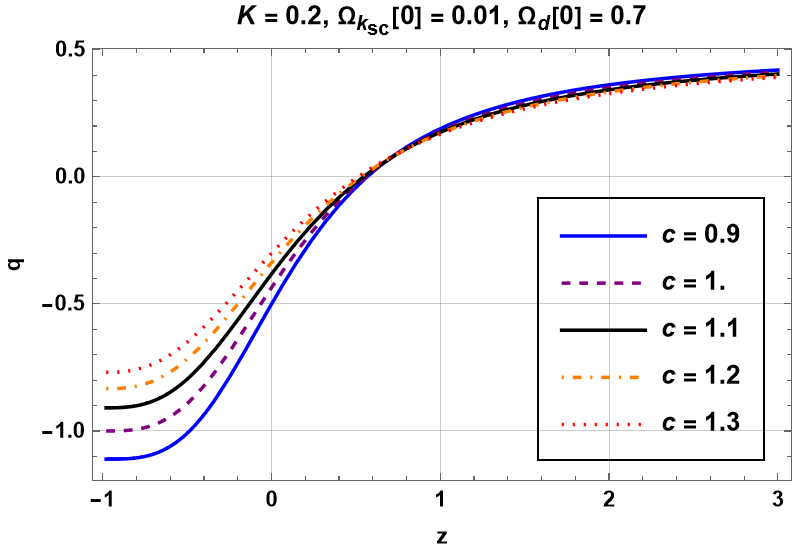}} 
\subfloat[]{\includegraphics[scale=0.4]{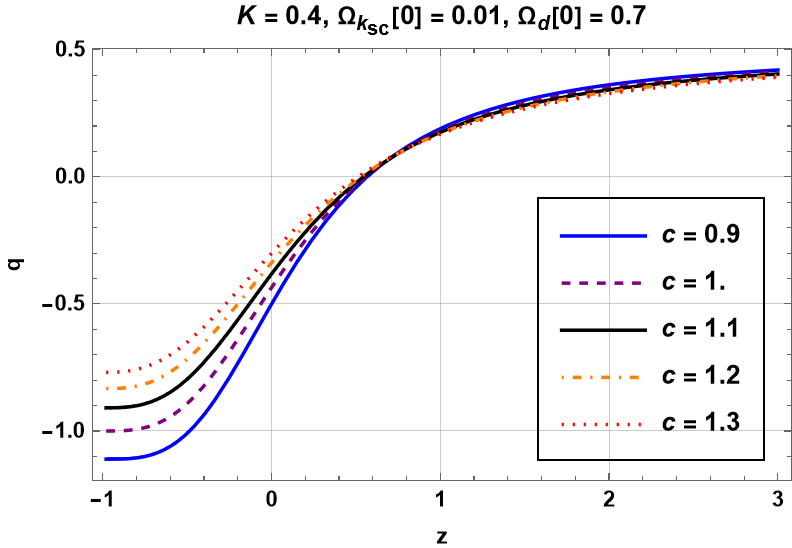}}
\subfloat[]{\includegraphics[scale=0.4]{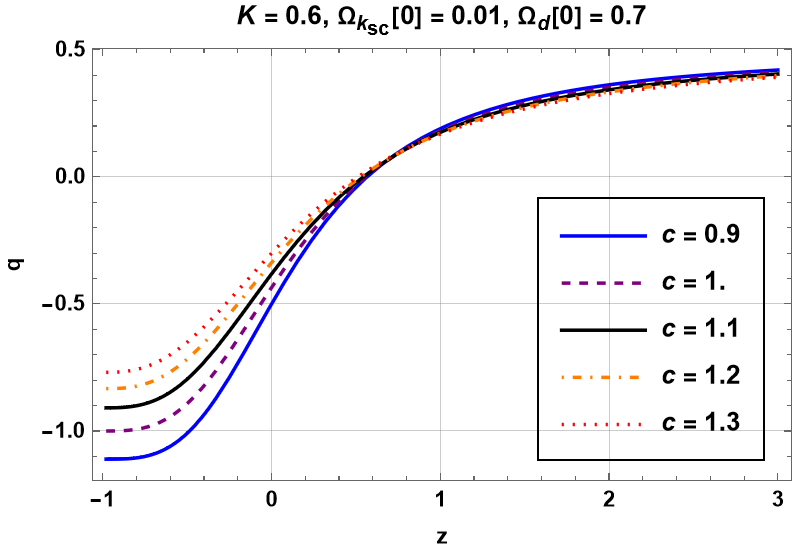}}
\caption{\begin{small}The variation in the deceleration  parameter $q$ with redshift $z$ for the closed universe,  in units when  $M_p^2=1, \quad H_0=67.9$. In order to be consistent with data, we now imposed $\Omega_d(x =- \ln(1 + z) = 0)= \Omega_d[0]= 0.7$.\end{small}}
\label{P5}
\end{figure}

\begin{figure}[H]
\subfloat[]{\includegraphics[scale=0.4]{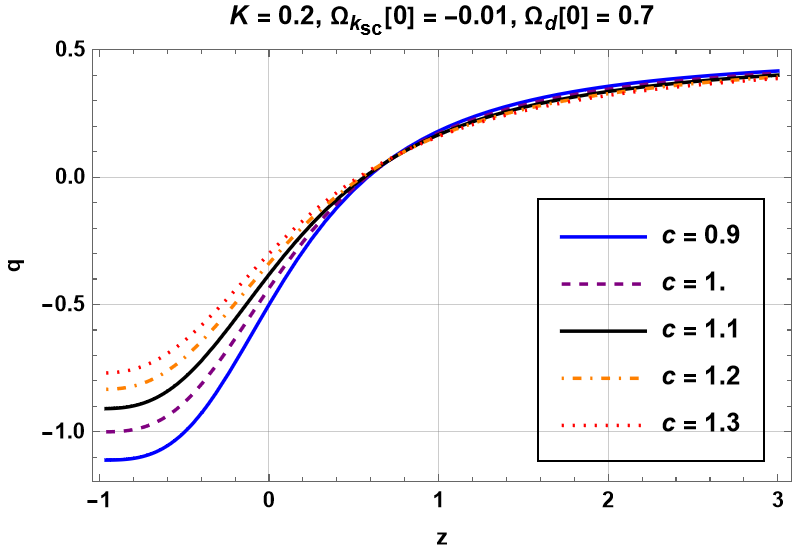}} 
\subfloat[]{\includegraphics[scale=0.4]{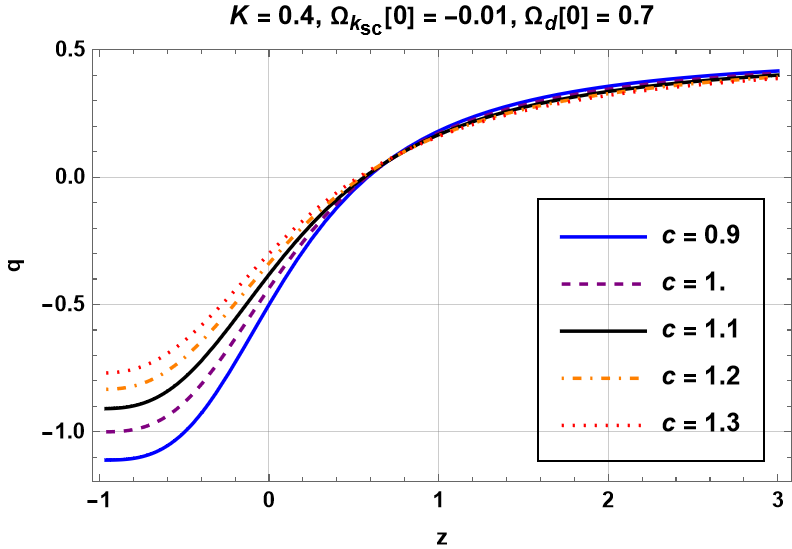}}
\subfloat[]{\includegraphics[scale=0.4]{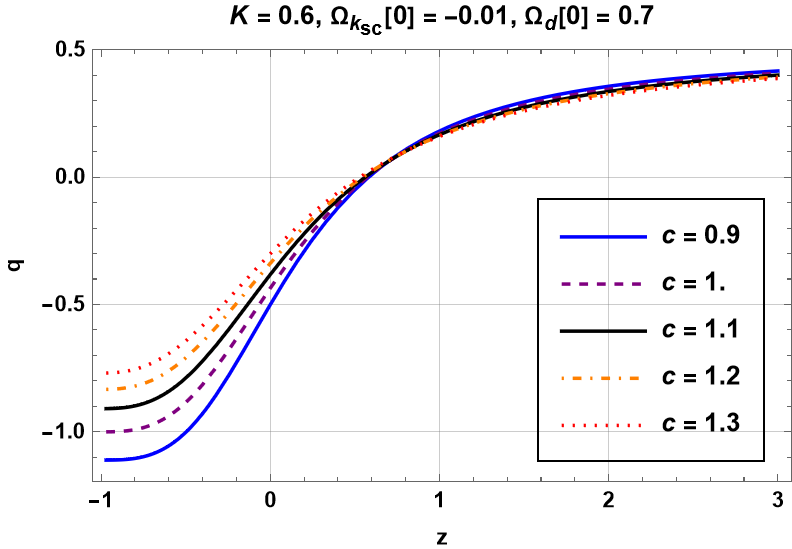}}
\caption{\begin{small}The variation in the deceleration parameter $q$ with redshift $z$ for the open universe,  in units when  $M_p^2=1, \quad H_0=67.9$. In order to be consistent with data, we now imposed $\Omega_d(x =- \ln(1 + z) = 0)= \Omega_d[0]= 0.7$.\end{small}}
\label{P6}
\end{figure}

Deceleration parameter plots w.r.t. $z$ are depicted in figures \ref{P5} and \ref{P6} for closed and  open universe models respectively by fixing the parameter $K$ and varying $c$ values. 
Both the list of figures clearly state that the universe started evolving in a decelerated phase. The phase shifted from deceleration to acceleration at about $z \approx 0.6$. The future of the universe will follow the accelerated phase irrespective of the nature of the model whether  closed or open.

\begin{figure}[H]
\subfloat[]{\includegraphics[scale=0.4]{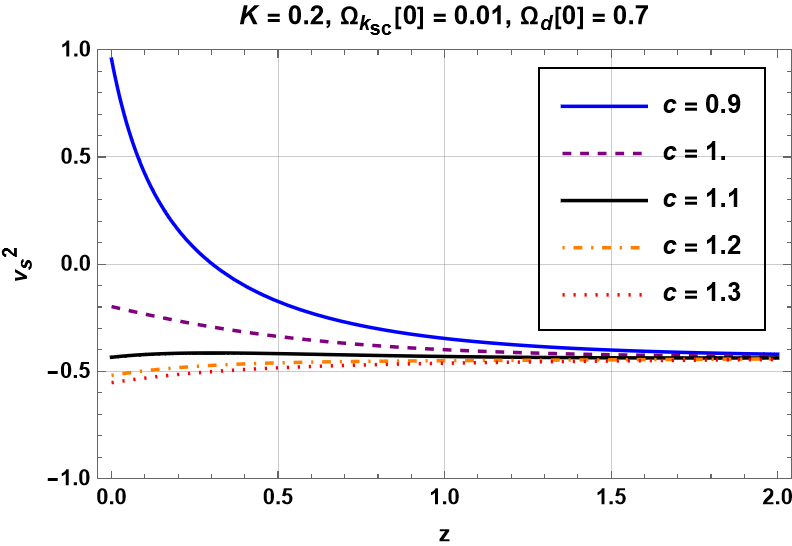}} 
\subfloat[]{\includegraphics[scale=0.4]{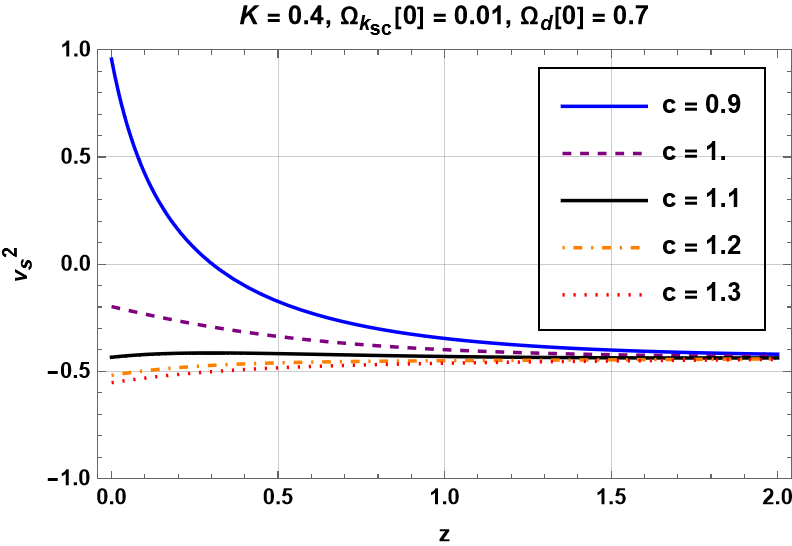}}
\subfloat[]{\includegraphics[scale=0.4]{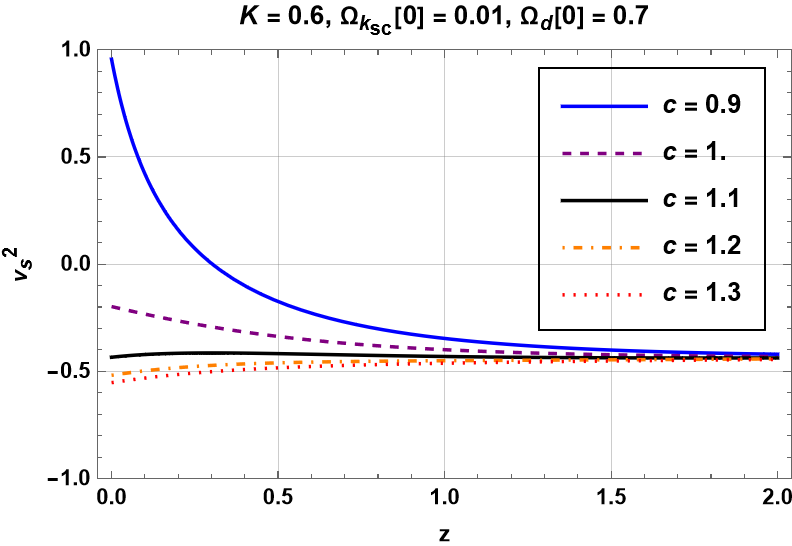}}
\caption{\begin{small}The variation in the squared sound speed  $v_s^2$ with redshift $z$ for the closed universe,  in units when  $M_p^2=1, \quad H_0=67.9$. In order to be consistent with data, we now imposed $\Omega_d(x =- \ln(1 + z) = 0)= \Omega_d[0]= 0.7$.\end{small}}
\label{P7}
\end{figure}

\begin{figure}[H]
\subfloat[]{\includegraphics[scale=0.4]{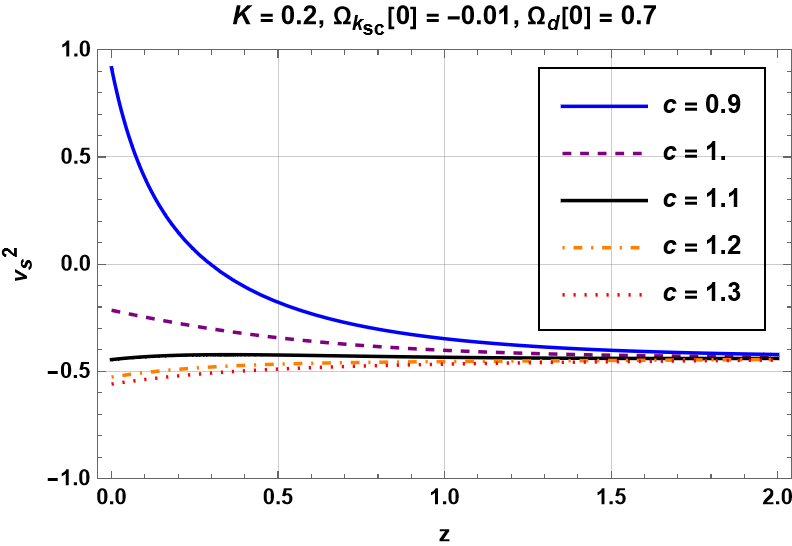}} 
\subfloat[]{\includegraphics[scale=0.4]{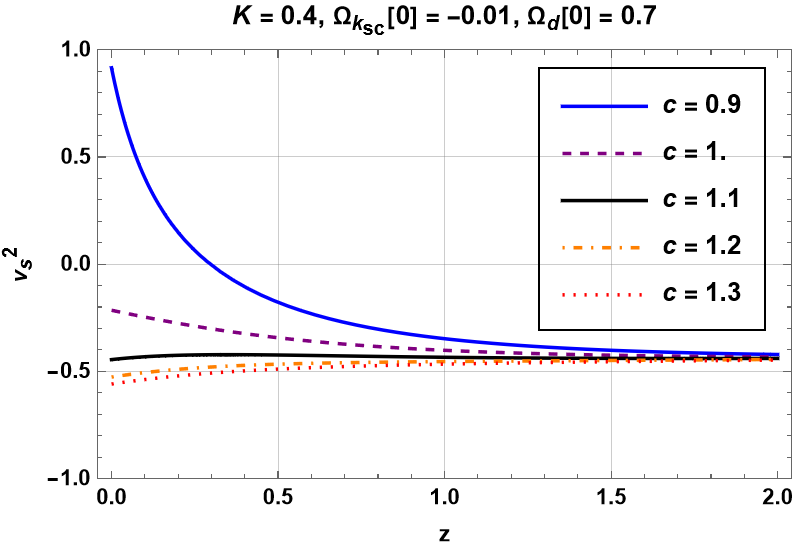}}
\subfloat[]{\includegraphics[scale=0.4]{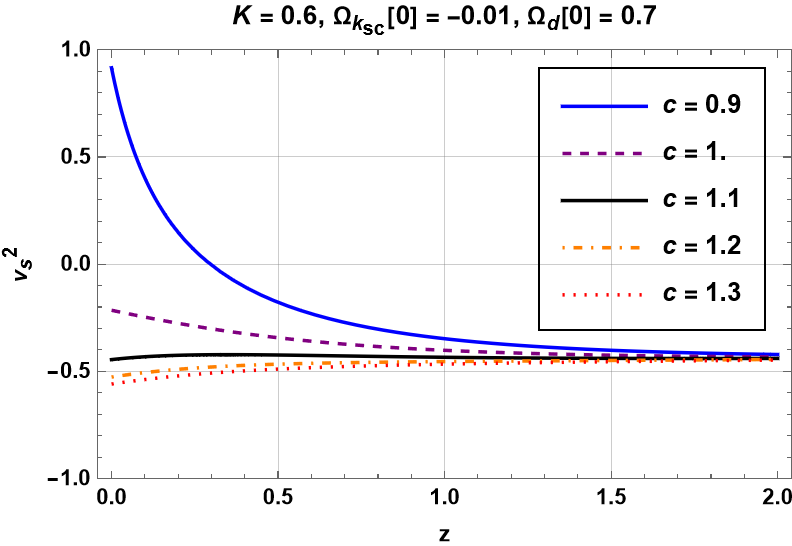}}
\caption{\begin{small} The variation in the squared sound speed  $v_s^2$ with redshift $z$ for the open universe,  in units when  $M_p^2=1, \quad H_0=67.9$. In order to be consistent with data, we now imposed $\Omega_d(x =- \ln(1 + z) = 0)= \Omega_d[0]= 0.7$.\end{small}}
\label{P8}
\end{figure}

The squared sound speed against redshift is displayed to assist  the model's conventional stability in figures \ref{P7} and \ref{P8} in the case of closed and open universe models respectively by varying $c$ values and fixed $K$. 
Both figures from \ref{P7} and \ref{P8} show that the KHDE model at hand is future stable for $c$ values less than $1$ otherwise unstable irrespective of $K$ values.

\section{Examination of Observed Data} 
This section discusses the non-flat KHDE model's advanced research analysis. A brief explanation is given of the datasets used in our research, with particular emphasis on how the Type Ia Supernova and CC (cosmic chronometer) mode were used to get the most recent Hubble parameter data (SNIa).

\subsection{Observationally Obtained Hubble Dataset}
\begin{figure}[H]
\subfloat[]{\includegraphics[scale=0.4]{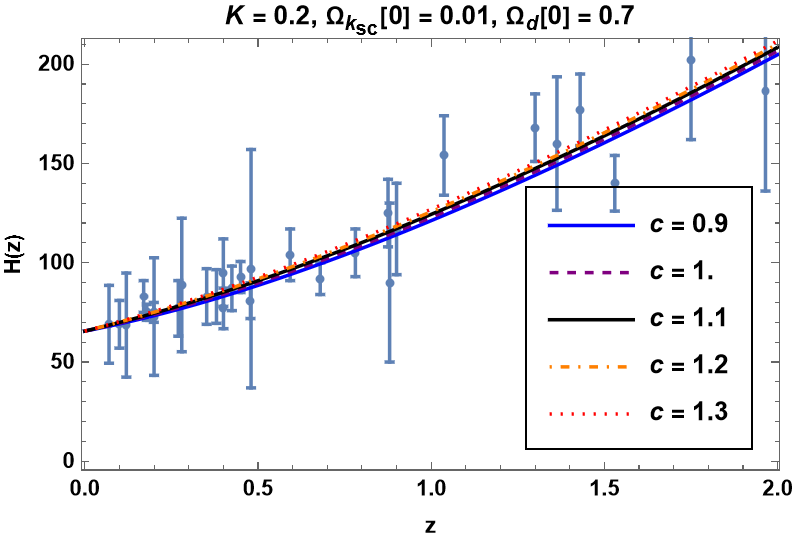}} 
\subfloat[]{\includegraphics[scale=0.4]{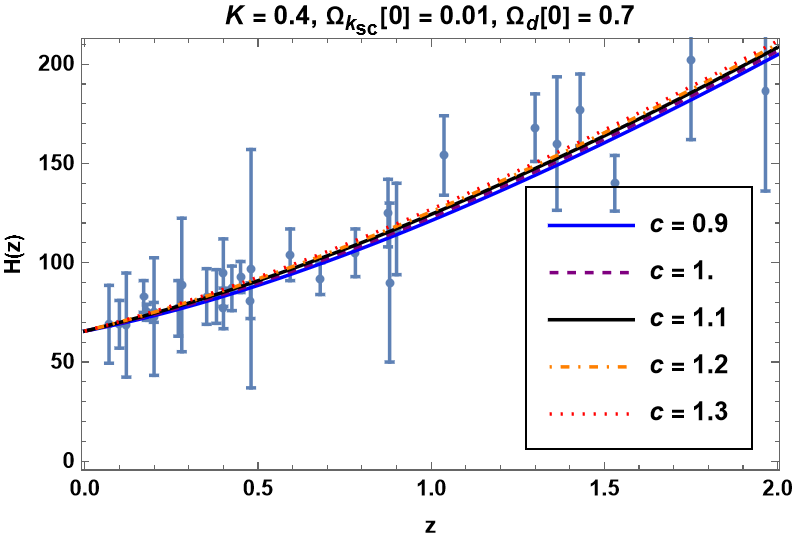}}
\subfloat[]{\includegraphics[scale=0.4]{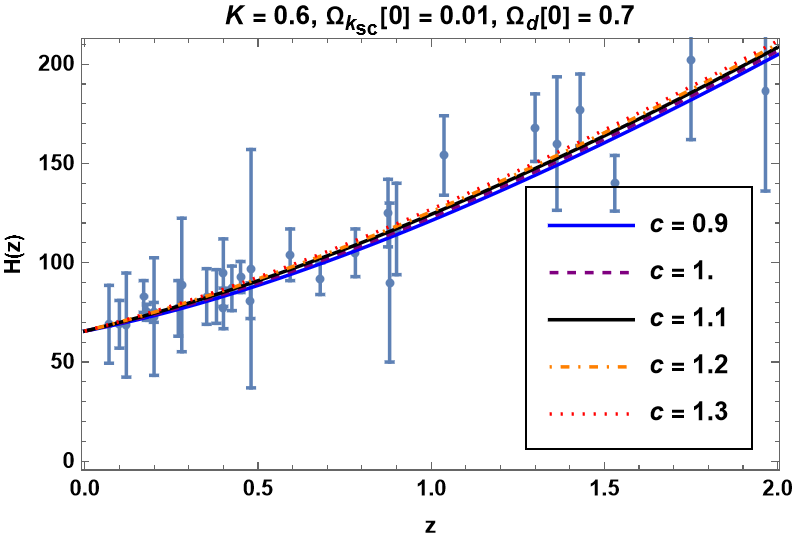}}
\caption{\begin{small}The variation in the Hubble parameter  $H$ with redshift $z$ for the closed universe,  in units when  $M_p^2=1, \quad H_0=67.9$. In order to be consistent with data, we now imposed $\Omega_d(x =- \ln(1 + z) = 0)= \Omega_d[0]= 0.7$. Bars represent the $30\; H(z)$ datapoints.\end{small}}
\label{P9}
\end{figure}

\begin{figure}[H]
\subfloat[]{\includegraphics[scale=0.4]{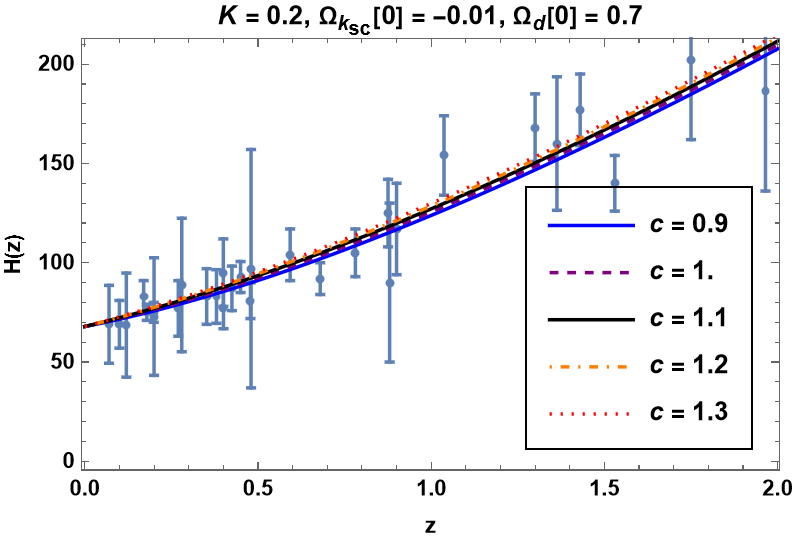}} 
\subfloat[]{\includegraphics[scale=0.4]{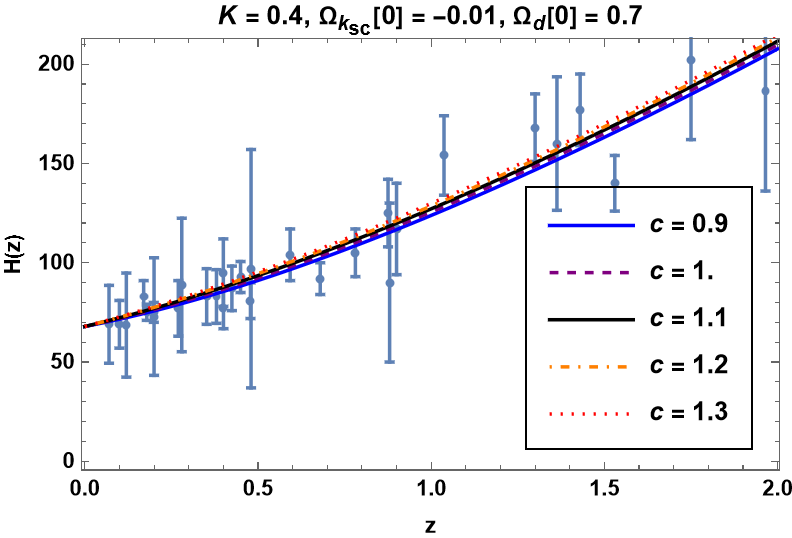}}
\subfloat[]{\includegraphics[scale=0.4]{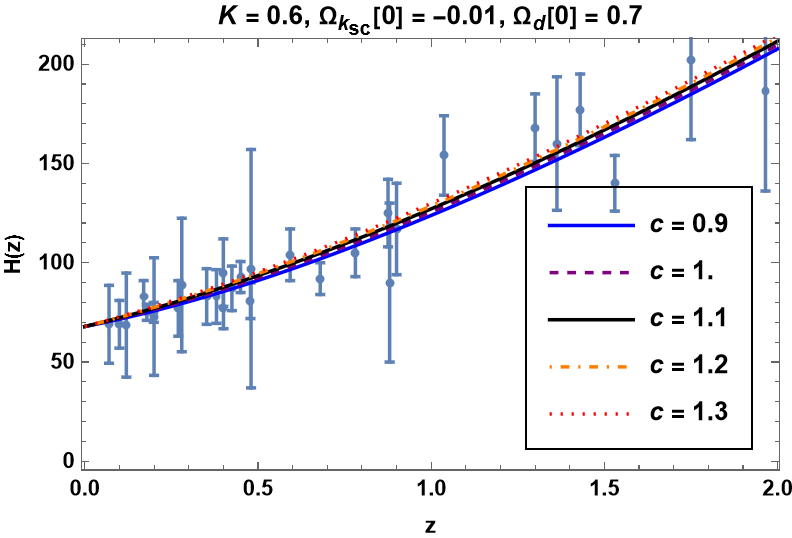}}
\caption{\begin{small} The variation in the Hubble parameter  $H$ with redshift $z$ for the open universe,  in units when  $M_p^2=1, \quad H_0=67.9$. In order to be consistent with data, we now imposed $\Omega_d(x =- \ln(1 + z) = 0)= \Omega_d[0]= 0.7$. Bars represent the $30\; H(z)$ datapoints.\end{small}}
\label{P10}
\end{figure}

We use the 30 H(z) observational dataset limitations from table 4 of \cite{Moresco16}, which covers the range of redshifts from $0.07$ to $1.965$. The cosmic chronometric (CC) method can be used to get this uncorrelated data. The reasoning behind collecting these data is based on the possibility that the Hubble dataset obtained via the CC approach is individualized. The method of different galaxy dating determines the CC data of the universe that is passively developing. Figures \ref{P9} and \ref{P10} demonstrate the development of $H(z)$ in the case of closed and open universe models respectively, and compares it to the said $30$  points of $H(z)$ data. Clearly, the current  model is completely consistent with the observationally obtained $H(z)$ dataset.

\subsection{Distance Modulus}
The observations taken from the SNIa are extremely helpful for studying cosmological models, especially as the main proof of an accelerating universe. As a result, in addition to the CC data, we also used a sample of $580$ Union 2.1 scores mixed with SNIa from the distance modulus dataset \cite{SupernovaCosmologyProject11}. The redshift-luminosity distance connection is a well-known observational method for studying the development of the universe \cite{Liddle2000}. The light travelling towards a faraway luminescent body is redshifted as a result of the universe's expansion, and this causes us to be able to calculate the $z$-depependent luminosity distance ($D_L$). We may calculate a source's flux using the luminosity distance, which is expressed by 
\begin{equation}
D_L = a_0r(1+z), \label{e32}
\end{equation}
with radial coordinate of the source $r$. Another definition of $D_L$ is given by \cite{Copeland06}
\begin{equation}
D_L=\dfrac{c_{light}(1+z)}{H_0}\int_{0}^{z} \dfrac{1}{E(z)}\mathrm{d}z, \label{e33}
\end{equation}
with $E(z)=\dfrac{H(z)}{H_0}$. And hence the distance modulus is given by 
\begin{equation}
\mu=25+5\log_{10}\left(\dfrac{D_L}{M_p}\right). \label{e34}
\end{equation}
Equation(\ref{e33}) yields the relation for the distance modulus $\mu$ given by
\begin{equation}
\mu =25+5\log_{10} \left[\dfrac{c_{light}(1+z)}{H_0}\int_{0}^{z} \dfrac{1}{E(z)}\mathrm{d}z\right]. \label{e35}
\end{equation}	
\\

\begin{figure}[H]
\subfloat[]{\includegraphics[scale=0.4]{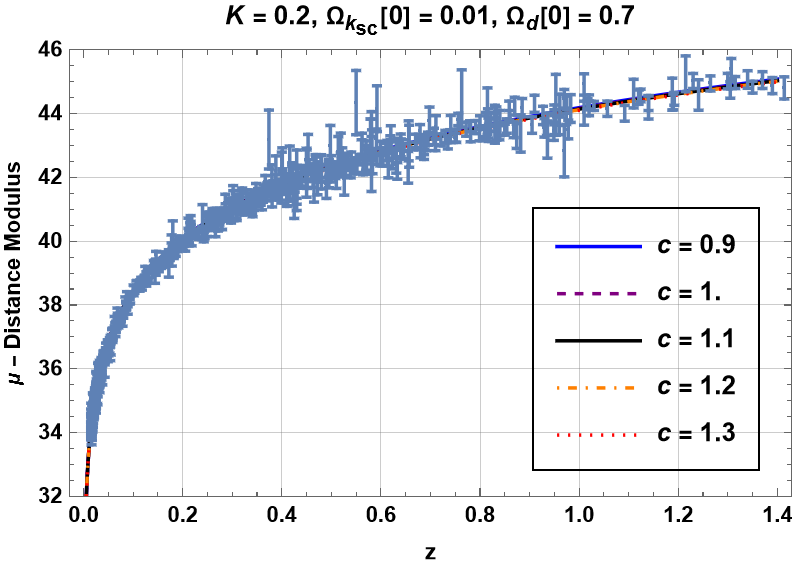}}
\subfloat[]{\includegraphics[scale=0.4]{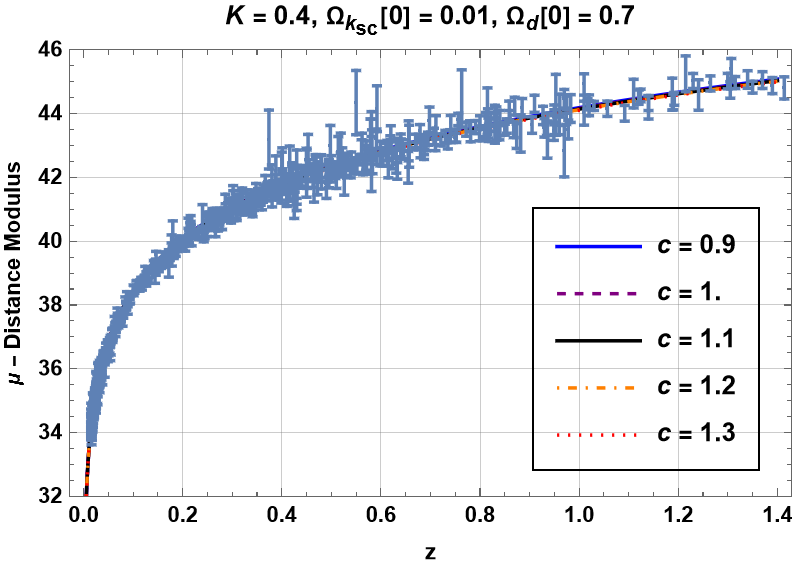}}
\subfloat[]{\includegraphics[scale=0.4]{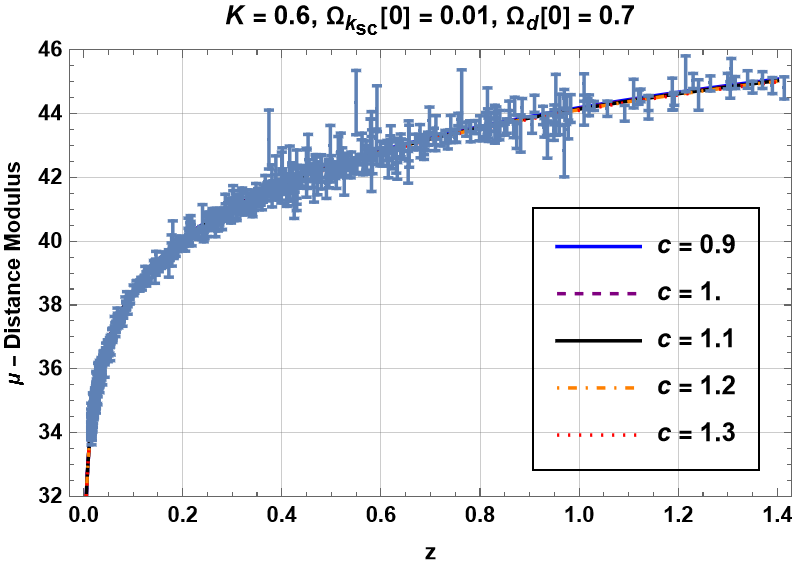}}
\caption{\begin{small} The variation in the luminosity distance $\mu$ with redshift $z$ for the closed universe,  in units when  $M_p^2=1, \quad H_0=67.9$. In order to be consistent with data, we now imposed $\Omega_d(x =- \ln(1 + z) = 0)= \Omega_d[0]= 0.7$. Bars display the $580$ Union 2.1 scores along with the SNIa results from the distance modulus data set. \end{small}}
\label{P11}
\end{figure}

\begin{figure}[H]
\subfloat[]{\includegraphics[scale=0.4]{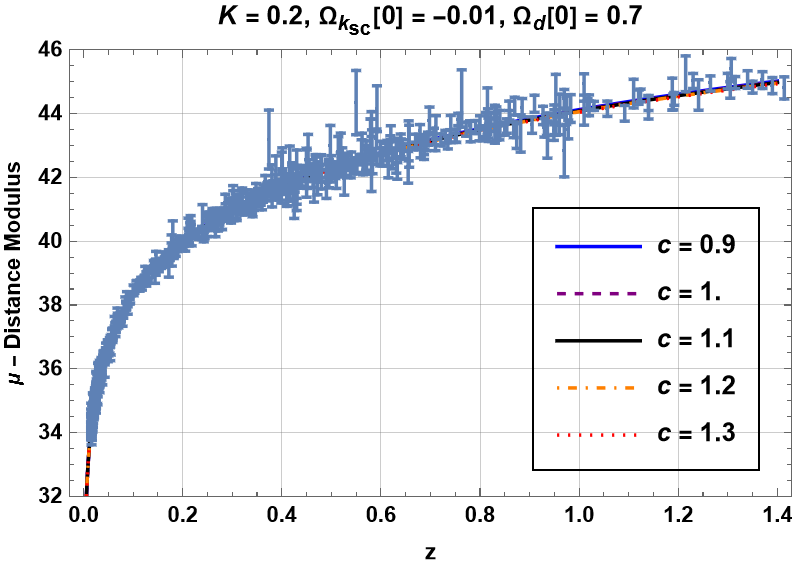}} 
\subfloat[]{\includegraphics[scale=0.4]{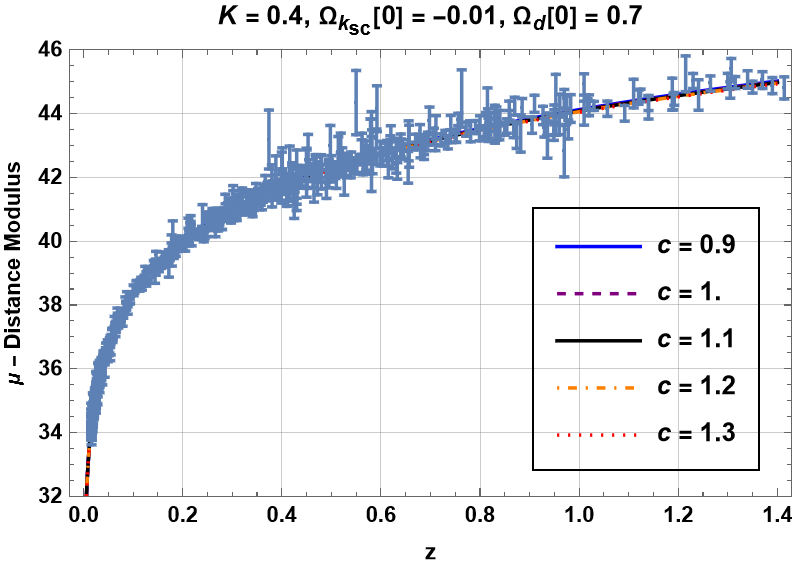}}
\subfloat[]{\includegraphics[scale=0.4]{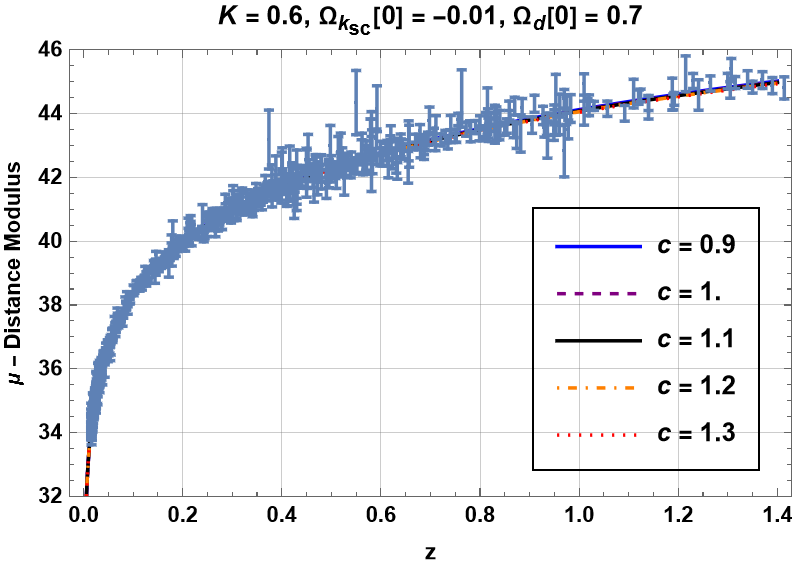}}
\caption{\begin{small} The variation in the luminosity distance $\mu$ with redshift $z$ for the open universe,  in units when  $M_p^2=1, \quad H_0=67.9$. In order to be consistent with data, we now imposed $\Omega_d(x =- \ln(1 + z) = 0)= \Omega_d[0]= 0.7$. Bars display the $580$ Union 2.1 scores along with the SNIa results from the distance modulus data set.\end{small}}
\label{P12}
\end{figure}

By combining the $580$ Union 2.1 \cite{SupernovaCosmologyProject11} scores and the SNIa data , figures \ref{P31} to \ref{P36} display the $z$-dependent period of distance modulus $\mu(z)$ for the non-flat KHDE model, appropriately. The error graph of the present model is presented by a solid line on togetherness of $580$ Union 2.1 results with SNIa datasets. Figures \ref{P11} and \ref{P12} show how the current model significantly reflects the observed $\mu(z)$ values for each data point.

\section{Summary of Results} 
In this paper, we built the KHDE with future event horizon as IR cutoff in a non-flat universe with model parameters $K$ and $c$. KHDE is constructed by applying Kaniadakis entropy to a cosmological framework instead of normal Bekenstein Hawking entropy and the holographic principle. We first offered a straightforward differential equation for the holographic dark energy density parameter $\Omega_d$ in order to examine the cosmic applications of KHDE. Additionally, as a function of $\Omega_d$, we deduced an analytical expression for the holographic dark energy equation-of-state parameter $w_d$ while accounting for both closed and open spatial geometry. Although the aforementioned differential equation can be analytically solved in an implicit form when $K=0$ and $\gamma=0$, in general case it can not be solved analytically, hence one must numerically elaborate it.\\

A rich behavior is also displayed by the associated dark energy EoS parameter, which can exhibit quintessence-like,  or quintom like experience by crossing  the phantom-divide  before or after the present. We then conducted a thorough examination, demonstrating that the KHDE model may properly reflect the universe's thermal history, including the sequence of matter and dark energy epochs. The change from deceleration to acceleration  occurs in the vicinity of  $z\approx 0.6$ in agreement with observations, before it leads to a complete dark energy dominance in the long term. The squared sound speed parameter behavior shows that the KHDE model is stable for the present epoch  but unstable for the very beginning epoch. Furthermore, we tested the scenario using observational data from the distance modulus dataset samples from $580$ Union 2.1 scores, and the combined SNIa data with 30 CC data points for $H(z)$, to validate the values of the parameters $K$ and $c$  under consideration, and found that the agreement is quite strong. \\

However, in order to more precisely constrain the new parameters, the present analysis should be expanded with more data from the BAO (Baryon Acoustic Oscillation), SNIa, and CMB probes. Further research on this topic is on the way, and further studies will likely disclose more insights.

\end{document}